\documentclass[journal]{IEEEtran}
\usepackage{fancyhdr}
\usepackage{amsmath} 
\usepackage{amssymb}
\usepackage{amsfonts}
\usepackage{graphicx}
\usepackage{algorithm}
\usepackage{algorithmic}
\usepackage{lettrine}
\usepackage{booktabs}
\usepackage{bbding}
\usepackage{threeparttable}
\ifCLASSINFOpdf
 
\else
 
\fi

\begin{document}
%
\title{Adaptive Worker Grouping for Communication-Efficient and Straggler-Tolerant Distributed SGD \thanks{Feng Zhu, Jingjing Zhang and Xin Wang are with the Department of Communication Science and Engineering, Fudan University, Shanghai 200433, China (e-mail: 20210720072@fudan.edu.cn; jingjingzhang@fudan.edu.cn; xwang11@fudan.edu.cn).} \thanks{Osvaldo Simeone is with the Department of Informatics, King's College London, London WC2R 2LS, U.K. (e-mail: osvaldo.simeone@kcl.ac.uk).}}

\author{Feng~Zhu,
        Jingjing~Zhang,
        Osvaldo~Simeone,~\IEEEmembership{Fellow,~IEEE,}
        and~Xin~Wang,~\IEEEmembership{Senior~Member,~IEEE}}

\maketitle
\thispagestyle{fancy}
\lhead{}
\chead{}
\rhead{}
\lfoot{}
\cfoot{}
\rfoot{\thepage} 
\renewcommand{\headrulewidth}{0pt} 
\renewcommand{\footrulewidth}{0pt} 
\pagestyle{fancy}
\rfoot{\thepage}
\begin{abstract}
Wall-clock convergence time and communication load are key performance metrics for the distributed implementation of stochastic gradient descent (SGD) in parameter server settings. Communication-adaptive distributed Adam (CADA) has been recently proposed as a way to reduce communication load via the adaptive selection of workers. CADA is subject to performance degradation in terms of wall-clock convergence time in the presence of stragglers. This paper proposes a novel scheme named grouping-based CADA (G-CADA) that retains the advantages of CADA in reducing the communication load, while increasing the robustness to stragglers at the cost of additional storage at the workers. G-CADA partitions the workers into groups of workers that are assigned the same data shards. Groups are scheduled adaptively at each iteration, and the server only waits for the fastest worker in each selected group. We provide analysis and experimental results to elaborate the significant gains on the wall-clock time, as well as communication load and computation load, of G-CADA over other benchmark schemes.
\end{abstract}


%
\IEEEpeerreviewmaketitle

\section{Introduction}
\lettrine[lines=2]{S}{tochastic} gradient descent (SGD)-based distributed learning has become an enabling technology for many artificial intelligence applications \cite{dean2012large}\cite{smith2017federated}. Wall-clock convergence time and communication load between workers and parameter servers (PS) are key performance indicators for distributed learning \cite{li2014scaling}\cite{ahmed2012scalable}\cite{ho2013more}. Wall-clock time performance is affected by workers that may be straggling \cite{li2014communication}, 
since in the standard implementation the PS needs to wait for all workers to respond at each iteration. Also, the communication overhead between the PS and the workers for the standard implementation grows linearly with the number of workers.

To address these issues, several techniques have been developed, including gradient coding (GC) and grouping \cite{tandon2017gradient}, which leverage storage and computation redundancy to mitigate the impact of stragglers, and adaptive selection, which selects workers adaptively to reduce the communication load \cite{chen2018lag}. This paper proposes for the first time to combine grouping with adaptive selection for the distributed implementation of SGD (see Fig. 1).

\begin{figure}[t]
\centering
\includegraphics[width=2.8in]{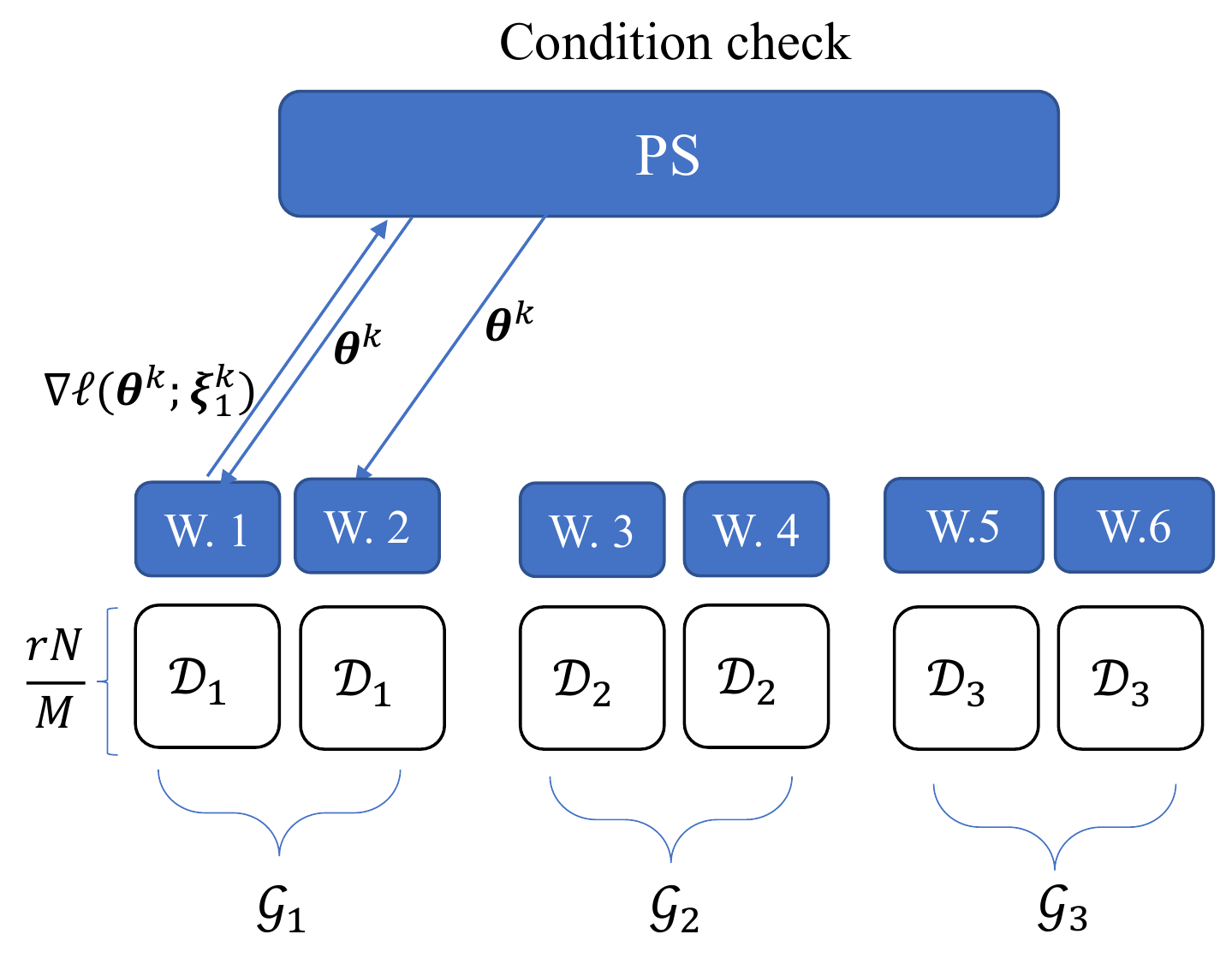}
\caption{Illustration of the proposed adaptive worker grouping technique with $M=6$ workers and $M_G=2$ workers per group, assuming $\mathcal{G}_1$ is the selected group.}
\end{figure}

\subsection{Related Work}
\subsubsection{Stochastic Gradient Descent (SGD)}
In large-scale machine learning, SGD has become the primary algorithm to trade convergence rate for computation complexity \cite{bottou2010large}. Many algorithms aiming to reduce the variance of SGD have been designed. Notable examples include the works in \cite{johnson2013accelerating}, \cite{roux2012stochastic} and \cite{shalev2013stochastic} that introduce stochastic variance reduced gradient (SVRG), stochastic average gradient (SAG) and stochastic dual coordinate ascent (SDCA), respectively. Also, adaptive SGD algorithms like AdaGrad \cite{duchi2011adaptive}, Adam \cite{kingma2014adam} and AMSGrad \cite{reddi2019convergence} have been demonstrated to be effective and efficient improvements of SGD for deep learning tasks.

\subsubsection{Gradient Coding and Grouping}
GC and grouping were introduced in \cite{tandon2017gradient} as means to speed up iterations in distributed gradient descent (GD) at the cost of storage and computation redundancy by allowing the PS to wait only for a subset of fastest workers. This work has been extended in several directions. In \cite{ozfatura2019gradient}, the authors developed algorithms to leverage partial computations at the stragglers; the communication and computation properties of GC were studied in \cite{li2015coded}, \cite{li2017fundamental} and \cite{ye2018communication}; while GC was extended to distributed SGD in \cite{wang2019erasurehead} and \cite{bitar2020stochastic}.

\subsubsection{Adaptive Selection}
Adaptive worker selection was introduced in \cite{chen2018lag} for distributed GD with the lazily aggregated gradient (LAG) scheme. The work \cite{zhang2020lagc} combined the idea of GC and LAG to develop lazily aggregated GC (LAGC), which achieves good results both in countering stragglers and in reducing communication load. The idea of LAG was extended to SGD via the lazily aggregated stochastic gradient (LASG) in \cite{chen2020lasg}; while Adam was substituted for the standard SGD in LASG 
to develop the communication-adaptive distributed Adam (CADA) in \cite{chen2021cada}.

\subsection{Main Contributions}
This paper proposes a novel straggler-tolerant and communication-efficient scheme for SGD-based distributed learning, which combines the advantages of adaptive selection and grouping. The main idea is to adaptively schedule groups of workers at each iteration and then wait only for the fastest worker in each selected group. The proposed scheme can be interpreted as a grouping-based version of CADA \cite{chen2021cada}, and is hence referred to as grouping-based CADA (G-CADA). It can also be viewed as a generalization of LAGC \cite{zhang2020lagc} from distributed GD to SGD. Importantly, unlike LAGC, G-CADA does not increase the computational load at the workers, since the mini-batch size does not depend on the storage redundancy. To gauge the performance of the proposed scheme, complexity analysis and numerical results are provided in terms of wall-clock time, as well as communication and computation loads.

The rest of this article is organized as follows. Section II presents the system model; the CADA scheme is reviewed in Section III; and Section IV introduces the proposed G-CADA scheme. Section V includes the analysis of different schemes in terms of different metrics; numerical results are given in Section VI and conclusions are drawn in Section VII.

\section{System Model}
\subsection{Setting}
The PS has available a global training dataset $\mathcal{D}=\{z_n=(\boldsymbol{x}_n,y_n)\}_{n=1}^N$, with $\boldsymbol{x}_n$ being a $d$-dimensional vector and $y_n$ being a scalar label, and the objective here is to address the empirical risk minimization problem
\begin{align}
\min _{\boldsymbol{\theta} \in \mathbb{R}^{p}} \mathcal{L}(\boldsymbol{\theta};\mathcal{D})=\frac{1}{N}\sum_{z \in \mathcal{D}} \ell(\boldsymbol{\theta};z)\label{problem}
\end{align}
for some smooth loss function $\ell(\boldsymbol{\theta};z)$. To this end, SGD is applied by the PS as
\begin{align}
    \boldsymbol{\theta}^{k+1}=\boldsymbol{\theta}^{k} - \alpha^k F^k(\hat{\boldsymbol{\nabla}}^k\mathcal{L}), \label{update}
\end{align}
where $\alpha^k$ is the stepsize, superscript $k$ denotes the iteration index, $\hat{\boldsymbol{\nabla}}^k\mathcal{L}$ is an estimate of the gradient $\nabla_{\boldsymbol{\theta}}\mathcal{L}(\boldsymbol{\theta}^k;\mathcal{D})$, and $F^k(\cdot)$ is some function that can be used to implement memory mechanisms such as Adam \cite{kingma2014adam}. The estimate $\hat{\boldsymbol{\nabla}}^k\mathcal{L}$ is obtained by leveraging parallel computing on multiple workers as discussed next.

Prior to the start of the iterations (\ref{update}), the PS distributes the dataset $\mathcal{D}$ among the set of $M$ workers in set $\mathcal{M}\triangleq\left \{  1,...,M\right \}$. Each worker $m\in\mathcal{M}$ is assigned a sub-data set $\mathcal{D}_m$ of $rN/M$ samples, where integer $r\geq 1$ is defined as the storage redundancy factor. In particular, $r>1$ implies that all data points are stored, and hence can be processed, at $r>1$ workers. Following the principle of adaptive selection \cite{chen2018lag}, at the beginning of each iteration $k$, the PS selects a subset of workers $\mathcal{M}_D^k\subseteq\mathcal{M}$, and it sends them the current global iterate $\boldsymbol{\theta}^k$. Each selected worker $m \in \mathcal{M}_D^k$ computes the local gradient $\nabla_{\boldsymbol{\theta}}\ell(\boldsymbol{\theta}^k; \boldsymbol{\xi}_m^k)=\nabla_{\boldsymbol{\theta}}(\sum_{z\in \boldsymbol{\xi}_m^k}\ell(\boldsymbol{\theta}^k; z))$ where $\boldsymbol{\xi}_m^k$ is a mini-batch of fixed size randomly selected from the dataset $\mathcal{D}_m$ at iteration $k$. 

Due to the fact that some workers might be straggling, only a subset $\mathcal{M}_U^k\subseteq\mathcal{M}_D^k$ of fastest workers that complete the computation upload the gradients. The PS then aggregates the uploaded gradients from the workers in subset $\mathcal{M}_U^k$ along with stale gradients from the workers in subset $\Tilde{\mathcal{M}}\triangleq\mathcal{M}\backslash\mathcal{M}_U^k$, and arrives at the estimated gradient
\begin{align}
\hat{\boldsymbol{\nabla}}^k\mathcal{L}=\sum_{m \in \mathcal{M}_U^k}\nabla\ell(\boldsymbol{\theta}^k; \boldsymbol{\xi}_m^k) + \sum_{m \in \Tilde{\mathcal{M}} }\nabla\ell(\boldsymbol{\theta}^{k-\tau_m^k}; \boldsymbol{\xi}_m^{k-\tau_m^k}), \label{aggregate}
\end{align}
 where $\tau_m^k\geq 1$ is the number of iteration elapsed since the last update from worker $m$, which we refer to age of information (AoI) of worker $m$ at iteration $k$. Finally, parameter $\boldsymbol{\theta}^k$ is updated through (\ref{update}). 
\subsection{Performance Metrics}
We explore the performance of different distributed SGD-based schemes in terms of wall-clock time, communication, and computation complexities as in \cite{zhang2020lagc}. To this end, for each iteration $k$, the computing time of each worker $m$, denoted by $T_m^k$, is assumed to be an exponential random variable with mean $\eta>0$. Note that the mini-batch size is fixed, and hence the distribution of time $T_m^k$ does not depend on the redundancy $r$, unlike in \cite{zhang2020lagc}. The variables $\{T_m^k\}_{m \in \mathcal{M}}$ are independently and identically distributed (i.i.d.) across all workers and iterations. Since the PS has to wait for the slowest worker in subset $\mathcal{M}_U^k$ that needs to upload the gradient, the wall-clock time complexity per iteration is given as
\begin{equation}
    \bar{T}=\mathbb{E}\left [  \max _{m \in \mathcal{M}_U^k}\left\{T_{m}^{k}\right\} \right ]. \label{time_gap}
\end{equation}
The communication load per iteration is defined as the average sum of the number of workers that download the global parameter from the PS and the number of workers uploading their fresh gradients, i.e., 
\begin{equation}
    \bar{C}=\mathbb{E}\left [ |\mathcal{M}_D^k|+|\mathcal{M}_U^k| \right ]. \label{comm_gap}
\end{equation}
Finally, the computation load per iteration is defined as the total number of mini-batch gradients computed at the workers, i.e.,
\begin{equation}
    \bar{P}=\mathbb{E}\left [ |\mathcal{M}_D^k|\cdot \mu \right ], \label{comp_gap}
\end{equation}
where the constant $\mu$ denotes the size of the mini-batch in terms of number of samples.

\section{CADA}
In this section, we review a close variant of CADA \cite{chen2021cada} that is modified here to fit the system model described in Section II, in which worker selection is carried out at the PS (and not at the workers as in \cite{chen2021cada}).

CADA assumes no computational redundancy, i.e., $r=1$, and it splits the general dataset $\mathcal{D}$ into $M$ equal-sized disjoint datasets $\mathcal{D}_1, ..., \mathcal{D}_M$, with $\mathcal{D}_m$ allocated to worker $m$. At each iteration $k$, the PS includes in the subset $\mathcal{M}_D^k$ worker $m$ that violates the following condition introduced in \cite{chen2020lasg}:
\begin{align}
L_m^2\left\|\boldsymbol{\theta}^{k}-\boldsymbol{\theta}^{k-\tau_{m}^{k}}\right\|^{2}
\leq c\sum_{d=1}^{D}\left\|\boldsymbol{\theta}^{k+1-d}-\boldsymbol{\theta}^{k-d}\right\|^{2} \label{cada-ps},
\end{align}
where $L_m$ is the smoothness constant of the local function $\ell(\boldsymbol{\theta};\mathcal{D}_m)=\frac{M}{rN}\sum_{z\in\mathcal{D}_m}\ell(\boldsymbol{\theta};\mathcal{D}_m)$ of each worker and $c>0$ is some constant; and $D$ is an integer $D\geq 1$. The left-hand side of (\ref{cada-ps}) estimates the change in the squared norm of the gradient at worker $m$; and the right-hand side represents the per-worker average contribution to the global iterate over the $D$ latest iterations. Additionally, a worker is included in subset $\mathcal{M}_D^k$ if its AoI is greater than or equal to $D$.

The parameter $\boldsymbol{\theta}^k$ is updated with the AMSGrad rule \cite{reddi2019convergence}, which uses the exponentially weighted stochastic gradient $\boldsymbol{h}^{k+1}$ as the direction of update and the weighted stochastic gradient magnitude vector ${\boldsymbol{v}}^{k+1}$ to adaptively control the stepsize. Specifically, the updated rule is
\begin{subequations}
\begin{align}
    &\boldsymbol{h}^{k+1}=\beta_1 \boldsymbol{h}^k+(1-\beta_1) \hat{\boldsymbol{\nabla}}^k\mathcal{L} \label{a}\\
    &\boldsymbol{v}^{k+1}=\beta_2\hat{\boldsymbol{v}}^k+(1-\beta_2)(\hat{\boldsymbol{\nabla}}^k\mathcal{L})^2 \label{b} \\
    &\boldsymbol{\theta}^{k+1}=\boldsymbol{\theta}^k-\alpha^k(\epsilon \boldsymbol{I}+\hat{\boldsymbol{V}}^{k+1})^{-\frac{1}{2}}\boldsymbol{h}^{k+1} \label{c} 
\end{align}
\end{subequations}
where $\beta_1\in(0,1)$ and $\beta_2\in(0,1)$ are momentum weights; $\hat{\boldsymbol{v}}^{k+1}\triangleq\max(\hat{\boldsymbol{v}}^k, \boldsymbol{v}^{k+1})$ is the element-wise maximum;  $\hat{\boldsymbol{V}}^{k+1}$ is a $p\times p$ diagonal matrix whose diagonal vector is $\hat{\boldsymbol{v}}^{k+1}$; $\boldsymbol{I}$ is a $p\times p$ identity matrix; $\epsilon>0$ is a small number; and the square operation in (\ref{b}) is element-wise.

\section{Grouping-Based CADA (G-CADA)}
In this section, we introduce the proposed G-CADA scheme. G-CADA leverages storage redundancy, i.e., $r>1$, via grouping in order to improve the robustness to stragglers of CADA, while still retaining CADA's benefits in terms of communication and computational loads.

In CADA, prior to training, the $M$ workers are divided into $G$ groups, $\mathcal{G}_1, ..., \mathcal{G}_G$, each with the same number of workers $M_G=M/G$. The global dataset $\mathcal{D}$ is split into equal-sized disjoint datasets $\mathcal{D}_1, ..., \mathcal{D}_G$, and each worker in group $\mathcal{G}_g$ is assigned $\mathcal{D}_g$. This implies a storage redundancy factor $r=M_G$ where $M_G$ is chosen such that $M_G\leq r$. For each group $\mathcal{G}_g$, we define the group-wise AoI $\tau_g^k$, which is maintained by the PS.

At each iteration $k$, the PS selects groups, rather than individual workers as in CADA. This is done by choosing the groups $\mathcal{G}_g$ that violate the condition
\begin{align}
L_g^2\left\|\boldsymbol{\theta}^{k}-\boldsymbol{\theta}^{k-\tau_{g}^{k}}\right\|^{2}
\leq c\sum_{d=1}^{D}\left\|\boldsymbol{\theta}^{k+1-d}-\boldsymbol{\theta}^{k-d}\right\|^{2}, \label{gas}
\end{align}
where $L_g$ is the smoothness constant of the local function $\ell(\boldsymbol{\theta};\mathcal{D}_g)=\frac{M}{rN}\sum_{z\in\mathcal{D}_g}\ell(\boldsymbol{\theta};\mathcal{D}_g)$ of group $\mathcal{G}_g$. The condition has a similar interpretation to (\ref{cada-ps}). We also include group $\mathcal{G}_g$ if the AoI $\tau_g^k$ is larger than $D$.

After determining the subset $\mathcal{G}_D^k$ of selected groups, the PS sends parameter $\boldsymbol{\theta}^k$ to all the workers in the selected groups, and each worker $m$ in $\mathcal{G}_D^k$ computes $\nabla\ell(\boldsymbol{\theta}^k; \boldsymbol{\xi}_m^k)$. The fastest worker $m$ in each selected group $\mathcal{G}_g$ uploads the computed gradient $\nabla\ell(\boldsymbol{\theta}^k; \boldsymbol{\xi}_g^k)=\nabla\ell(\boldsymbol{\theta}^k; \boldsymbol{\xi}_m^k)$ to the PS.

Then the PS updates the AoI of the selected groups as $\tau_g^{k+1}=1$, while for other groups it sets $\tau_g^{k+1}=\tau_g^k$. Finally, the parameter $\boldsymbol{\theta}^k$ is updated via (\ref{a})-(\ref{c}). Similarly we define $\Tilde{\mathcal{G}}\triangleq\mathcal{G} \backslash \mathcal{G}_D^{k}$. The aggregated gradient is hence given as\\ $\hat{\boldsymbol{\nabla}}^k\mathcal{L}=\sum_{g \in \Tilde{\mathcal{G}}} \nabla \ell(\boldsymbol{\theta}^{k-\tau_{g}^{k}} ; \boldsymbol{\xi}_{g}^{k-\tau_g^k})+\sum_{g \in \mathcal{G}_D^{k}} \nabla \ell(\boldsymbol{\theta}^{k} ; \boldsymbol{\xi}_{g}^k)$. \\The complete procedure of G-CADA is summarized in Algorithm 1.


As detailed in Algorithm 1, while in CADA the PS has to wait for the slowest selected worker, in G-CADA the PS only needs to wait for the fastest worker in each selected group, which can potentially reduce the wall-clock time.

\begin{algorithm}[t]
\caption{G-CADA}
\label{alg1}
\begin{algorithmic}[1]
\REQUIRE number of groups $G=M/M_G$, stepsize $\alpha^k>0$, delay counter$\{\tau_g^0\}$, constants $\{c_d\}$, max delay $D$, smoothness constants $\{L_g\}$
\ENSURE $\boldsymbol{\theta}^0$, $k=0$
\REPEAT
\STATE the PS checks the condition (\ref{gas})
\FOR{each group $\mathcal{G}_g$ in $\mathcal{G}_D^k$ or satisfies $\tau_g^k\geq D$ in parallel}
\STATE all the workers in group $\mathcal{G}_g$ download $\boldsymbol{\theta}^k$ from the PS  
\STATE each worker $m$ in group $\mathcal{G}_g$ computes $\nabla \ell\left(\boldsymbol{\theta}^{k} ; \boldsymbol{\xi}_{m}^{k}\right)$ 
\STATE the fastest worker uploads the gradient $\nabla \ell\left(\boldsymbol{\theta}^{k} ; \boldsymbol{\xi}_{g}^{k}\right)$ with $\nabla \ell\left(\boldsymbol{\theta}^{k} ; \boldsymbol{\xi}_{g}^{k}\right)=\nabla \ell\left(\boldsymbol{\theta}^{k} ; \boldsymbol{\xi}_{m}^{k}\right)$
\ENDFOR
\STATE server updates $\{\boldsymbol{h}^k, \boldsymbol{v}^k\}$ and $\boldsymbol{\theta}^k$ via (\ref{a})-(\ref{c}) and sets $\{\tau_g^{k+1}=1\}_{g\in \mathcal{G}_D^k}$ and $\{\tau_g^{k+1}=\tau_g^k+1\}_{g\in \Tilde{\mathcal{G}}}$
\STATE $k=k+1$
\UNTIL convergence criterion is satisfied
\end{algorithmic}
\end{algorithm}

\section{Analysis}
In this section we analyze the wall-clock time complexity, communication complexity, and computation complexity of distributed Adam (a direct distributed implementation of Adam \cite{kingma2014adam}), CADA (as described in Section III) and G-CADA. Based on \cite{chen2020lasg}, we can conclude that all schemes have sublinear convergence rate for convex loss functions. Motivated by this, in this section, we analyze the per-iteration metrics defined in Section II.
\subsection{Wall-Clock Time}
Let $T_{a:b}$ be the $a$th order statistics of i.i.d. variables $\{T_i\}_{i=1}^b$, which is the $a$th smallest value in the set $\{T_i\}_{i=1}^b$, and $T_{a:a}:=T_a$. We have the average \cite{mallick2019rateless}
\begin{align}
\bar{T}_{a:b}=\mathbb{E}[T_{a:b}]=\eta (H_b-H_{b-a}), \label{exp_time}
\end{align}
where $H_a=\sum_{k=1}^a 1/k$ is the $a$th harmonic number.

\textit{Distributed Adam:} In distributed Adam, at each iteration, the PS broadcasts the parameter to all the workers and each worker $m$ computes the gradient with the current iterate $\nabla \ell(\boldsymbol{\theta}^{k} ; \boldsymbol{\xi}_{m}^k)$ and uploads it to the PS. Since the PS waits for all the workers, the average runtime per iteration is 
\begin{align}
\bar{T}_{D-Adam}=\bar{T}_M. \label{time_sgd}
\end{align}

\textit{CADA:} For CADA, from \cite{zhang2020lagc} and \cite{chen2018lag}, we can upper bound the average workers selected per iteration as
\begin{align}
\bar{M}=M\sum_{d=0}^D\frac{h(d)}{d+1}\leq M, \label{average_workers}
\end{align}
where we have defined the function $h(d)=(1/M)\sum_{m \in \mathcal{M}} \mathbb{I}\left(\bar{L}_{ d+1}^{2}<L_{m}^{2}<\bar{L}_{d}^{2}\right)$, where $\mathbb{I}$ is the indicator function, with $\bar{L}_{d}^{2}=c_d/(dM^2)
$ and $\bar{L}_0=\bar{L}_{D+1}=0$. Therefore, since the PS waits for all the selected workers, the average runtime per iteration is 
\begin{align}
\bar{T}_{CADA}=\bar{T}_{\bar{M}}. \label{time_cada}
\end{align}

\textit{G-CADA:} In G-CADA, the PS waits for the fastest worker in each selected group. Therefore the runtime $T_g^G$ for group $\mathcal{G}_g$ is the first order statistic of the random variables $\{T_i\}_{i\in \mathcal{G}_g}$. Let $\bar{T}_{a:G}^G$ be the average of the $a$th smallest number of the random variables $\{T_g^G\}_{g=1}^G$ and $\bar{T}_{G:G}^G=\bar{T}_{G}^G$. This can be evaluated as
\begin{align}
\bar{T}_{a:M_G}^G=\int_{0}^{+\infty}\left(1-\left(F^{G}(x)\right)^{a}\right)dx, \label{time_g} 
\end{align}
where the cumulative distribution function (CDF) of each variable $T_g^G$ is $F^{G}(x)=\sum_{j=1}^{M_{G}}{M_G \choose j}(F(x))^{j}(1-F(x))^{M_{G}-j}$. Defining the function $h_G(d)=(1/G)\sum_{g \in[G]} \mathbb{I}(\bar{L}_{G, d+1}^{2}<L_{g}^{2}<\bar{L}_{G, d}^{2})$, with $\bar{L}_{G, d}^{2}=c_d/(dG^2)
$ and $\bar{L}_0=\bar{L}_{D+1}=0$, the average groups selected per iteration can be upper bounded as
\begin{align}
\bar{G}=G\sum_{d=0}^D\frac{h_G(d)}{d+1}\leq G. \label{average_groups}
\end{align}
Since the PS waits for the slowest group, the average runtime per iteration for G-CADA is 
\begin{align}
\bar{T}_{G-CADA}=\bar{T}_{\bar{G}}^G.
\label{time_ps}
\end{align}
By comparing (\ref{time_ps}) with (\ref{time_sgd}) and (\ref{time_cada}), we observe that wall-clock time advantage is achieved as compared to distributed Adam and CADA. 
\subsection{Communication Load}
\textit{Distributed Adam:} The communication load per iteration for the distributed Adam scheme is 
\begin{align}
\bar{C}_{D-Adam}=\mathbb{E}\left [ |\mathcal{M}_D^k|+|\mathcal{M}_U^k| \right ]=2M, \label{comm_sgd}
\end{align}
since all the workers download the parameters and upload the gradients.

\textit{CADA:} For CADA, the communication load per iteration is 
\begin{align}
\bar{C}_{CADA}=\mathbb{E}\left [ |\mathcal{M}_D^k|+|\mathcal{M}_U^k| \right ]\leq 2\bar{M}, \label{comm_cada}
\end{align}
with $\bar{M}$ defined in (\ref{average_workers}), since only the selected workers download the parameter and upload the gradients.

\textit{G-CADA:} The communication load per iteration for G-CADA is
\begin{align}
\bar{C}_{G-CADA}=\mathbb{E}\left [ |\mathcal{M}_D^k|+|\mathcal{M}_U^k| \right ]\leq \bar{G}(M_G+1), \label{comm_gas}
\end{align}
since all the workers in the selected groups download the parameters and the fastest ones upload the gradient.

\subsection{Computation Load}
\textit{Distributed Adam:} The per-iteration computation load of the distributed Adam scheme is
\begin{align}
P_{D-Adam}=\mathbb{E}\left [ |\mathcal{M}_D^k|\cdot \mu \right ] = \mu M \label{comp_sgd}
\end{align}
since all the workers need to compute the gradient.

\textit{CADA:} The computation load per iteration for CADA is 
\begin{align}
P_{CADA}=\mathbb{E}\left [ |\mathcal{M}_D^k|\cdot \mu \right ] = \mu \bar{M} \label{comp_cada}    
\end{align}
since only the selected workers need to compute the gradients.

\textit{G-CADA:} For G-CADA, the computation load per iteration is 
\begin{align}
P_{G-CADA:}=\mathbb{E}\left [ |\mathcal{M}_D^k|\cdot \mu \right ] \leq \mu \bar{G}\cdot M_G \label{comp_wk}
\end{align}
since only the workers in the selected groups need to compute the gradients.

\section{Numerical Results}

In this section, we provide numerical results to compare the performance of the considered schemes in terms of training loss, communication load, and computation load with respect to wall-clock time. For comparison, we also consider distributed SGD, which applies standard constant-stepsize SGD. We consider the linear regression model with MNIST dataset and quadratic error loss function. We set a total of $M=12$ workers, number of groups $G=3$, $M_G=r=4$,  $\beta_1=0.9, \beta_2=0.999$ and $\mu=10^{-4}$ sec. The learning rate $\alpha^k$ in (\ref{update}) for distributed SGD is 2.6, while it is set to 0.01 for all other schemes. The constant $c$ in (\ref{cada-ps}) for CADA is $c=2$ and for G-CADA we set $c=0.3$ in (\ref{gas}).

Fig. 2 illustrates the loss function, communication complexity (\ref{comm_gap}) and computation complexity (\ref{comp_gap}) as a function of the wall-clock time. Comparing CADA with distributed Adam we observe that adaptive selection reduces communication and computation loads. G-CADA achieves further improvement in wall-clock time and communication load with respect to CADA thanks to grouping, while maintaining the same level of computation load as CADA when measured at the same value of training loss. For instance, given a training loss level of $10^{-1}$, G-CADA requires a wall-clock time of 0.068 sec, a communication load of 12,292, and a computation load of 9,219. In contrast, CADA requires a wall-clock time of 0.380 sec, a communication load of 19,292, and a computation load of 9,646.
\begin{figure}[t]
\centering
\includegraphics[width=3in]{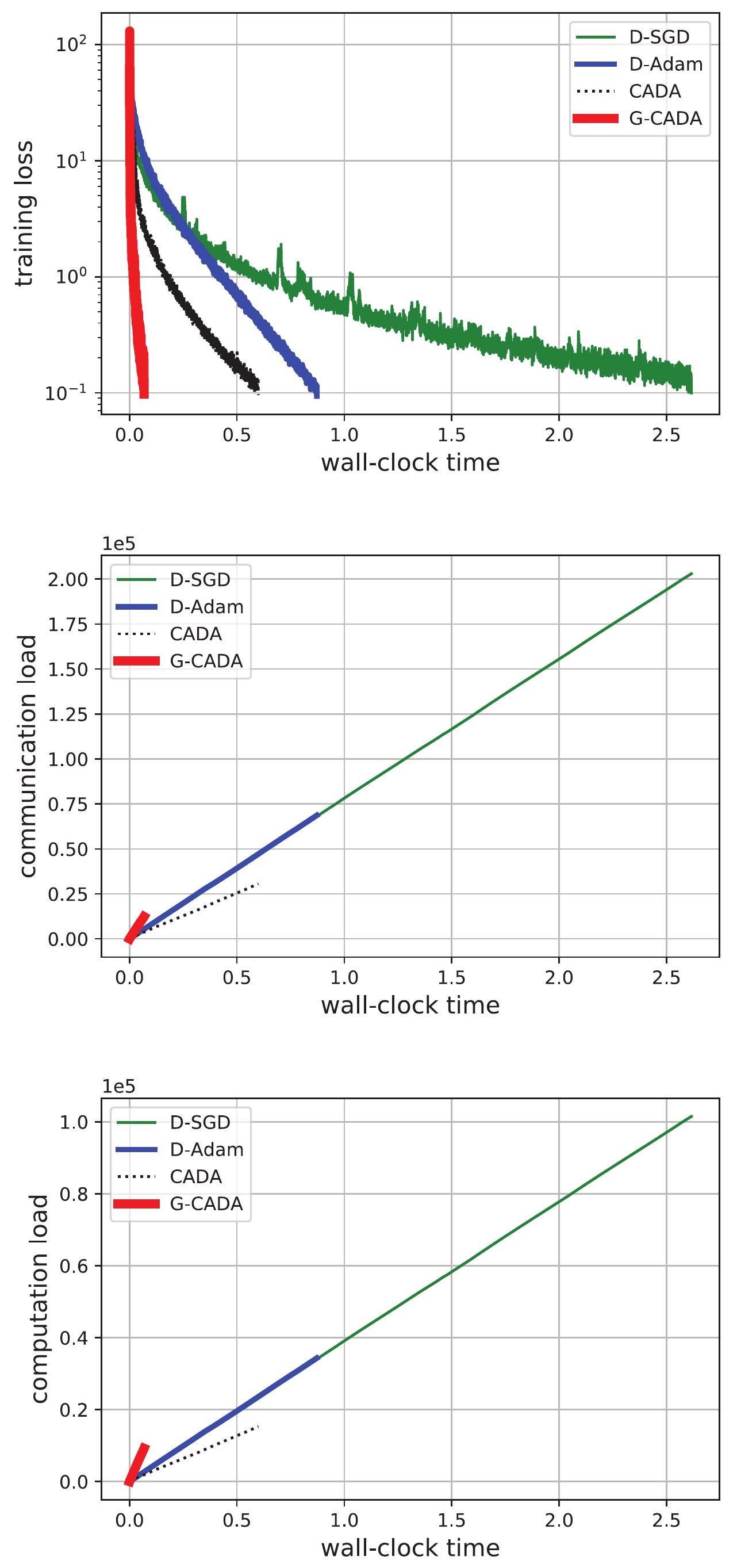}
\caption{Training loss, communication load (\ref{comm_gap}) and computation load (\ref{comp_gap}) against wall-clock time with exponential distribution for the computing times where $M=12, G=3, M_G=4$ and $\mu=10^{-4}$ sec.}
\end{figure}
\section{Conclusions}
In this paper, we have proposed a method that trades storage redundancy for wall-clock time and communication load in SGD-based distributed learning. The novel scheme, G-CADA, integrates grouping and adaptive selection. Grouping brings robustness to stragglers, while adaptive selection renders the system communication-efficient and decreases the computation load. Numerical results have shown that G-CADA achieves significant improvements in wall-clock time and reduced communication load, at the cost of storage redundancy.
\section{Acknowledgement}
The work of Feng Zhu and Jingjing Zhang has been supported by National Natural Science Foundation of China Grant No. 62101134. Osvaldo Simeone has received funding from the European Research Council (ERC) under the European Union's Horizon 2020 Research and Innovation Programme (Grant Agreement No. 725731). Xin Wang has been supported by the Innovation Program of Shanghai Municipal Science and Technology Commission Grant 20JC1416400, and the National Natural Science Foundation of China Grant No. 62071126.

\end{document}